\documentclass{ws-procs961x669}            

\usepackage{graphicx}
\usepackage{float}
\usepackage{epstopdf,cancel}
\usepackage{epsf,latexsym,bbm}
\usepackage[mathscr]{euscript}

\usepackage{amssymb,amsmath}
\usepackage{mathtools} 
\usepackage{times,graphics}
\usepackage{soul,xcolor}
 \usepackage{epsfig,ulem}
\def\6{\langle}
\def\9{\rangle}
\newcommand{\pad}{\partial}
\newcommand{\be}{\begin{equation}}
\newcommand{\ee}{\end{equation}}
\newcommand{\ba}{\begin{eqnarray}}
\newcommand{\ea}{\end{eqnarray}}

\def\cO{\mathcal{O}}
\newcommand{\hbb}{\mathbf{\hat b}}
\newcommand{\hbk}{\mathbf{\hat k}}
\newcommand{\hbf}{\mathbf{\hat f}}

\def\hbx{\hat{\mathbf{x}}}
\def\hby{\hat{\mathbf{y}}}
\def\hbz{\hat{\mathbf{z}}}

\newcommand\bk{{\bf{k}}}
\newcommand{\bu}{\mathbf{u}}
\newcommand{\bv}{\mathbf{v}}
\def\bof{\mathbf{f}}
\def\bv{\mathbf{v}}
\def\bu{\mathbf{u}}

\def\af{\mathsf{f}}
\def\ak{{\mathsf{k}}}
\def\aK{{\mathsf{K}}}
\def\an{{\mathsf{n}}}
\def\am{{\mathsf{m}}}
\def\aw{\mathsf{w}}

\def\fin{\mathrm{fin}}

\newcommand{\Omegab}{\mbox{\boldmath$\Omega$}}
\newcommand{\omegab}{\mbox{\boldmath$\omega$}}

\newcommand\sg{\textsl{g}}

\newcommand\vb{{\vec{b}}}

\newcommand\vk{{\vec{k}}}

\newcommand\vx{{\vec{x}}}

\newcommand\eR{{\mathscr{R}}}

\newcommand\rS{{\mathrm{S}}}

\newcommand{\hzeb}{\mbox{\boldmath$\hat\zeta$}}

\newcommand{\etal}{\textit{et al.}}

\newcommand{\defeq}{\vcentcolon=}
\newcommand{\eqdef}{=\vcentcolon}

\usepackage{url,hyperref}
\hypersetup{colorlinks,linkcolor={blue!55!black},citecolor={red!45!black},urlcolor={blue!45!black},breaklinks=true}

\begin{document}

\title{Light rays in the Solar system experiments: phases and displacements}

\author{Pravin Kumar Dahal$^*$ and Daniel R. Terno}

\address{Department of Physics and Astronomy, Macquarie University,\\
Sydney, NSW 2109, Australia\\
$^*$E-mail: pravin-kumar.dahal@hdr.mq.edu.au}

\begin{abstract}

Geometric optics approximation is sufficient to describe the effects in the near-Earth environment. In this framework  Faraday rotation is purely a reference frame (gauge) effect. However, it cannot be simply dismissed. Establishing local reference frame with respect to some distant stars leads to the Faraday phase error between the ground station and the spacecraft of the order of $10^{-10}$ in the leading post-Newtonian expansion of the Earth's gravitational field. While the Wigner phase of special relativity is of the order $10^{-4}$--$10^{-5}$. Both types of errors can be simultaneously mitigated by simple encoding procedures. We also present briefly the covariant formulation of geometric optic correction up to the subleading order approximation, which is necessary for the propagation of electromagnetic/ gravitational waves of large but finite frequencies. We use this formalism to obtain a closed form of the polarization dependent correction of the light ray trajectory in the leading order in a weak spherically symmetric gravitational field.

\end{abstract}

\keywords{Geometric optics; Wigner phase; Gravitational Faraday rotation; Post-Newtonian expansion; Gravitational spin Hall effect.}

\bodymatter

\section{Introduction}

Space deployment of quantum technology \cite{micius-1,time,qsat} brings it into
a weakly relativistic regime. As an unintended but fortunate  side effect,  low-Earth  orbit (LEO) quantum communication satellites   provide  new opportunities to test  fundamental physics. Once the tiny putative  physical effects fall within the
sensitivity range of these devices, they may impose constraints on practical quantum
communications, time-keeping, or remote sensing tasks\cite{pt:04,cqg,qs:12}. A more futuristic technology, such as the proposed solar gravity telescope\cite{Landis(2016),tt:18,tt:20} actually needs the general-relativistic effects
for its operation.

For flying qubits that are implemented as polarization states of photons\cite{photon} the dominant  source of relativistic errors in this setting is   the Wigner rotation (or phase), an effect special relativity (SR)\cite{pt:04,qs:12}.  Gravitational polarization rotation, also known as the gravitational Faraday effect \cite{gfaraday-1,gfaraday-2} occurs in a variety of astrophysical systems, such as accretion disks around astrophysical black hole candidates \cite{f-bh} or gravitational lensing \cite{f-gl}. This effect was the subject  of a large number of theoretical investigations, primarily within the geometric optics approximation \cite{gfaraday-1,gfaraday-2,f-gl,fl:82,tg-pol,C:92}, and also from the perspective of quantum communications\cite{fara-q}.  Interpretation of these results was until recently sometimes contradictory, as it is important to carefully analyze the relation between the reference frames of the emitter and the detector. Moreover, even if at the leading  order the gravitationally induced polarization rotation in the near-Earth environment  is pure gauge effect,  it cannot be simply dismissed. We provide a simple estimation of this emitter- and observer-dependent phase and give its explicit form in several settings.

Already in the leading  post-eikonal order  trajectories of light beams are affected by polarization.\cite{M:74,GBM:07,FS:11}. The optical gravitational spin Hall effect has  recently received a comprehensive treatment in Refs.~\citenum{OJDRPA:20,F:20}. Using this formalism we obtain a closed form of the leading correction in case of a weak spherically symmetric gravitational field. As expected, on the scale of the Solar system --- be it the near-Earth environment or a focal plane of the solar gravity telescope at 600a.u.--- the effect is negligible. However, it is interesting conceptually and its scaling indicates that it may play a much more important role in the strong gravity regions\cite{F:20}.

The rest of this article is organized as follows. In Sec.~\ref{wigner} we review the SR effects. Polarization rotation in general stationary spacetimes is described in Sec.~\ref{fr3}, where we also evaluate the effects in communication with Earth-orbiting satellites.  In Sec.~\ref{post-G} we briefly summarize the main techniques of Refs.~\citenum{OJDRPA:20,F:20} and then obtain the polarization-dependent changes in the light ray trajectories in the solar system experiments.

 We work with $c=\hbar=G=1$. The  constants $G$ and $c$ are restored in a small number of expressions where their presence is helpful. The spacetime metric $\sg_{\mu\nu}$ has a signature $-+++$. The four-vectors are distinguished by the sans font, $\ak$, $k^\mu=(\ak)^\mu$. The three-dimensional spatial metric is denoted as $\gamma_{mn}$, and three-dimensional vectors are set in boldface, $\bk$, or are referred to by their explicit coordinate form, $k^m$. The inner product in the metric $\gamma$ is denoted as $\bk\!\cdot\!\bof$, and the unit vectors in this metric are distinguished by the caret, $\hbk$, $\hbk\cdot\hbk\equiv1$. Post-Newtonian calculations employ a fiducial Euclidean space. Euclidean vectors are distinguished by arrows, $\vk$. Components of the two types of vectors may coincide,  $(\bv)^m=(\vb)^m$, but $\vb\!\cdot\!\vk=\sum_{k=1}^3 v^m k^m$. Accordingly, the coordinate distance is the Euclidean length of the radius vector, $r\equiv\sqrt{\vx\!\cdot\!\vx}$. 

 \begin{figure}[htbp]
\includegraphics[width=0.50\textwidth]{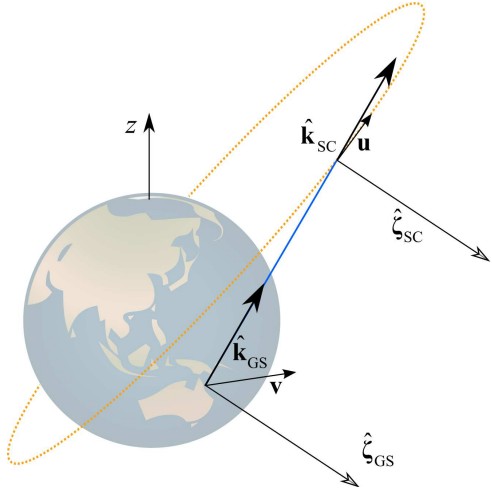}
\caption{ \label{fig1} Scheme of the communication round between the ground station (GS) and the spacecraft (SC). The light ray is highlighted in blue.
 All the vectors are indicated in the Earth-centered inertial frame. In the flat spacetime approximation the
 unit tangent vectors to the ray $\hbk_{\mathrm{GS}}$ and $\hbk_{\mathrm{SC}}$ at the GS and the SC, as well as the direction to an infinitely distant guide star, $\hzeb_{\mathrm{GS}}$ and $\hzeb_{\mathrm{SC}}$, respectively coincide.  Velocity of the GS at the emission of the signal is $\bv$ and velocity of the SC at the moment of detection is $\bu$.  }
\end{figure}

\section{Wigner rotation and special relativistic effects} \label{wigner}

Consider one round of communications between the ground station (GS) and   a low energy orbit spacecraft (SC). The problem is most conveniently analyzed in the geocentric system, the origin of which coincides with the centre of the Earth at the moment of emission. If we direct the $z$-axis of the system along the Earth's angular momentum, then the velocity $\bv$ of the GS lies in the xy-plane. Whereas, the velocity $\bu$ of the SC and the initial propagation direction $\hbk$ are arbitrary when expressed in the global frame. This setting is depicted in  Fig.~\ref{fig1}.

Quantum states of a photon with a definite four-momentum $\ak=(|\bk|,\bk)$ can be represented either as Hilbert space vectors or as complex polarization vectors in the usual three-dimensional space,
\be
|\Psi_\ak\9=f_+|\ak,+\9+f_-|\ak,-\9 \quad \Leftrightarrow \quad \hbf_\bk=f_+{\hbb}_\hbk^++f_-{\hbb}_\hbk^-,
\ee
with the transversal vectors $\hbb^\pm_\hbk$, $\bk\cdot\hbb^\pm_\hbk=0$. The correspondence is rooted in the relationship between   finite-dimensional and unitary representations of the Poincar\'{e} group \cite{wig,wkt}.

Unitary operators $U$ that describe the state transformation between the Lorentz frames are obtained via the induced representation of the Poincar\'{e} group. Basis states that correspond to an arbitrary momentum $(\ak)^\mu\equiv k^\mu$ are defined with the help of standard Lorentz transformation $L(k^\mu)$, that takes the four momentum from the standard value $k_\rS^\mu$ to $k^\mu$. For massless particles, $k_\rS^\mu= (1,0,0,1)$ and
\begin{equation}
    L(k^\mu)= R(\hbk) B_z(\xi_{|\bk|}),
\end{equation}
where $R(\hbk)= R_z(\phi) R_y(\theta)$ rotates the $z$-axis into the direction $\hbk$ by performing rotations around the $y$- and $z$-axis by angles $\theta$ and $\phi$, respectively. These rotations follow the boost $B_z(\xi_{|\bk|})$ along the $z$-axis, that brings the magnitude of momentum to $|\bk|$.

The states of an arbitrary momentum are defined via
\be
 |\ak,\pm\9\defeq U\big(L(\ak)\big)|\ak_\rS,\pm\9,
\ee
while the standard right- and left-circular polarization vectors are defined as
 \be
 \hbb_\hbk^\pm\defeq R(\hbk)(\hat{\mathbf{x}}\pm i \hat{\mathbf{y}})/ \sqrt{2},
 \ee
where the linear polarization vectors are $\hbb_1\defeq R(\hbk)\hbx$ and $\hbb_2\defeq R(\hbk)\hby$, respectively. Alternatively, these vectors can be obtained as
 \be
 \hbb_2=\frac{\hat{\mathbf{z}} \times \hat{\mathbf{k}}}{|\hbz\times\hbk|}, \qquad \hbb_1=\hbb_2\times\hbk. \label{b12k}
 \ee
 The explicit form of the polarization  four-vector  $\af_\ak$,  $\af_\ak\cdot \ak=0$ depends on the gauge
 \cite{pt:04,fara-q,hks:85}.

Under arbitrary Lorentz transformation, states transform (apart from the normalization factor) via
\begin{equation}
   U(\Lambda) |\ak,\pm\rangle= U(\Lambda L(\ak)) |\ak_\rS,\pm\rangle= U(L(\Lambda \ak)) U(W) |\ak_s,\pm\rangle,
\end{equation}
where the   transformation $W= L^{-1}(\Lambda \ak) \Lambda L(\ak)$ to the subgroup (Wigner's little group) group that leaves $\ak_s$ invariant,
\begin{equation}
    U(W) |\ak_\rS,\sigma\rangle= \sum_{\pm'} D(W)_{\sigma\sigma'} |\ak_\rS,\sigma'\rangle.
\end{equation}
The matrices $D(W)_{\sigma\sigma'}$ form  the representation of the little group. For massless particles an element of  the little group $W$ can be decomposed as
\begin{equation}
    W= S R_z(\varpi),
\end{equation}
where $S$ is a translation in the $(xy)$-plane and $R_z(\varpi)$  a rotation. As   translations do not contribute to the physical degrees of freedom of photons, the state  $|\ak,\pm\rangle$ transforms as
\begin{equation}
    U(\Lambda) |\ak,\pm\rangle= e^{\pm i\varpi} |\Lambda\ak,\pm\rangle.
\end{equation}

There are no generic explicit  expressions for $\varpi$. Their evaluation is  not considerably simpler if $\Lambda=\eR$, where $\eR$ is a rotation (as there is no risk of confusion we use the same designation for the four-dimensional matrices of spatial dimensions and for their $3\times 3$ blocks). However, as the transformation law of $\hbf$ can be obtained from the three-dimensional form of the Lorentz transformations of the transversal electromagnetic wave, in this case \cite{pt:04,lpt-1}
\be
U(\eR)|\Psi_\ak\9  \Leftrightarrow \eR \hbf_\bk.
\ee
Moreover, an arbitrary  rotation around the  direction $\hbb_2$, $R_{\hbb_2}(\alpha)$,
does not introduce a phase $\varpi$ \cite{tg-pol}. This provides the motivation for introduction of the so-called Newton gauge that we review below.

In   communications with the Earth-orbiting satellites settings of Fig.~\ref{fig1} the Wigner phase is the dominant relativistic effect\cite{DT:21}. While the Wigner phase is of the order $10^{-4}$--$10^{-5}$, we will see below that establishing the local reference frame with respect to some distant stars leads to the Faraday phase error of the order of $10^{-10}$.

 \section{Faraday rotation and general relativistic effects} \label{fr3}

Here we describe the effects of gravity on polarization in the geometric optics approximation, wave it can be considered a vector that is simply affected by the geodesic  motion of null particles to which it is attached. We present the gravitational Faraday effect in a way that clearly separates the gauge-independent  part from the effects of the reference frame choices\cite{tg-pol} and is convenient for the near-Earth calculations that use the post-Newtonian approximation.  Then we demonstrate that at the leading order the polarization rotation is a purely gauge effect and evaluate it for a practically useful choice of the reference frames.

The equation of geometric optics are obtained by performing the short wave expansion of the wave equation in the Lorentz gauge\cite{mtw,H:19},
\begin{equation}
    A^\mu=a^\mu e^{i {\psi}},\label{a8}
\end{equation}
where $a^\mu$ is the slowly varying complex amplitude and ${\psi}$ is the rapidly varying real phase. In later calculations, we use the wave vector $k_\mu\defeq{\psi}_{,\mu}$, the squared amplitude $a=\left(  a^{\mu*} a_\mu\right)^{1/2}$ and the polarization vector $f^\mu=a^\mu/a$ is transverse to the trajectory, $f^\mu k_\mu=0$.

The eikonal equation
\be
\sg^{\mu\nu}\frac{\pad\psi}{\pad x^\mu}\frac{\pad\psi}{\pad x^\nu}=0, \label{HJequation}
\ee
 is the leading term in  the expansion of the wave equation \cite{mtw,H:19}  is the Hamilton-Jacobi equation for a free massless particle on a given background  {spacetime}. It  allows description of   light propagation in terms of fictitious massless particles.

The wave vector $k_\mu$, which is normal to the hypersurface of constant phase is null $k^\mu k_\mu=0$, and is geodesic
\begin{equation}
    k^\mu \nabla_\mu k^\nu=0,
\end{equation}
as it is the gradient of a scalar function. Similarly, the polarization vector $f^\mu$  is parallel propagated along it
\begin{equation}
   k^\mu \nabla_\mu f^\nu=0.
\end{equation}

 A convenient three dimensional representation of the evolution of polarization vectors is possible in stationary spacetimes. Static observers follow the congruence of timelike Killing geodesics that define projection from spacetime manifold $\cal M$ onto the three dimensional space $\Sigma_3$, $\Pi:{\cal M}\to \Sigma_3$. The metric $\sg_{\mu\nu}$ on $\cal M$ can be written in terms of three dimensional scalar $h$, a vector $\mathbf g$ with component $\sg_m$ and a metric $\gamma_{m n}$ on $\Sigma_3$ as
\begin{equation}
    d\textsl{s}^2= -h \left(dx^0- \sg_m dx^m\right)^2+ dl^2,
\end{equation}
where $h=-\sg_{0 0}$, $g_m= -\sg_{0 m}/\sg_{0 0}$, and the three dimensional distance $dl^2= \gamma_{m n} dx^m dx^n$, where
\begin{equation}
    \gamma_{m n}= \sg_{m n}- \frac{\sg_{0 m} \sg_{0 n}}{\sg_{0 0}}.
\end{equation}
Using the relationships between the three and four dimensional covariant derivatives, the propagation equations in a stationary spacetimes result in the following three dimensional equations
\begin{equation}
    \frac{D\hbk}{Ds}= \Omegab \times \hbk, \qquad \frac{D\hbf}{Ds}= \Omegab \times \hbf, \label{pt15}
\end{equation}
where $D/Ds$ is the covariant derivative in three dimensional space with metric $\gamma_{m n}$. Thus, both the polarization and propagation vectors are rigidly rotated with an angular velocity
\begin{equation}
    \Omegab= 2 \omegab- (\omegab.\hbk)\hbk- \mathbf E_g\times \hbk,
\end{equation}
where $\omegab$ and $\mathbf E_g$ could be interpreted as the gravitoelectric and gravitomagnetic field respectively
\begin{equation}
    \omegab= -\frac{1}{2} k_0 \nabla\times \mathbf g,\qquad \mathbf E_g= -\frac{\nabla h}{2 h}.
\end{equation}

In flat spacetimes, polarization basis is uniquely fixed by the Wigner little group construction. However, in general curved background, Wigner construction must be performed at every point. This is because, in the absence of a global reference direction, the standard polarization triad $(\hbb_1, \hbb_2, \hbk)$ is different at every locations. Given such choice, the net polarization can be found by starting with the initial polarization $f_{\fin}(x_\mathrm{in})$, parallel propagating it according to the rule Eq.~\eqref{pt15} and then using the decomposition
\begin{equation}
    f_{\fin}(x_{\fin})= \cos\chi \hbb_1+ \sin\chi \hbb_2,
\end{equation}
to read off an angle. For Schwarzschild spacetime, $\omegab=0$, and thus differentiation of this equation gives
\begin{equation}
    \frac{d\chi}{ds}= \frac{1}{\hbf\cdot \hbb_1}\hbf\cdot\frac{D\hbb_2}{Ds}. \label{ph19}
\end{equation}
This equation implies that polarization rotation is a pure gauge effect. The phase remains zero if the standard directions are set with the help of the local free fall acceleration $\mathbf{w}$ of a stationary observer. At each point in the spacetime we choose the direction of the standard reference momentum, or equivalently the $z$-axis of our standard polarization triad, to be $\hzeb\defeq\mathbf{w}$.  For a photon with momentum $\mathbf{k}$ we choose the linear polarization vector $\hat{\mathbf{b}}_2$ to  point in the direction $\hzeb \times \hat{\mathbf{k}}$, and finally we choose $\hat{\mathbf{b}}_1\defeq \hbb_2\times\hbk$ such that it completes the orthonormal triad $( \hat{\mathbf{b}}_1, \hat{\mathbf{b}}_2, \hat{\mathbf{k}})$. This construction is known as the Newton gauge \cite{tg-pol}. With this convention
$\Omegab=-\mathbf E_g\times\bk\equiv\Omega \hbb_2$ and thus $\chi\equiv 0$ along the trajectory. However, such choice of standard polarization direction is practically unfeasible. We will see below the consequences of setting the $z$-axis with the help of a guide star.

Electromagnetic radiation and massless particles are not affected by Newtonian gravity. The post-Newtonian expansion\cite{will} is conveniently organized in powers of $\epsilon^2\sim GM/c^2\ell\sim v^2/c^2$, where $-GM/\ell$ is the (maximal) typical potential and $v$ is a typical velocity of massive particles. The parameter $\epsilon$ helps to keep  track of the orders and is set to unity at the end of the calculations.
The leading post-Newtonian contributions are of order $\epsilon^2$; to take gravitomagnetic effects into account, we need contributions up to $\epsilon^3$.

The post-Newtonian expansion of the metric near a single slowly rotating quasirigid gravitating body, up to $\epsilon^3$, assuming that the underlying theory of gravity is general relativity is
\begin{equation}
    d\textsl{s}^2= -V^2(r) c^2 dt^2+ \vec{R}\! \cdot\! d\vec{x} c dt+ W^2(r) d\vec{x}\!\cdot\! d\vec{x},
\end{equation}
where
\begin{equation}
    V(r)=1-\epsilon^2 \frac{U}{c^2}, \qquad W(r)= 1+ \epsilon^2 \frac{U}{c^2}.
\end{equation}
The Newtonian gravitational potential $-U=$ $-G M Q(r,\theta)/r$ $\simeq-G M/r$  depends on the mass $M$ and the higher order multipoles. The frame-dragging term is
\begin{equation}
    \vec{R}= -\epsilon^3 \frac{4 G}{c^3} \frac{\vec J \times \vec x}{r^3},
\end{equation}
where $\vec J$ is the angular momentum of the rotating body. Hence, we see that the gauge invariant polarization rotation is absent in the leading order post-Newtonian expansion and the Faraday phase at order $\epsilon^2$ is a reference frame effect. We again revert to the units $G=c=1$. To obtain leading order contributions to the phase and polarization, photon trajectories only need to be expanded up to $\epsilon^2$
\begin{equation}
    \vec k= \vec n- \epsilon^2 \frac{2 M}{r(t)} \vec n- \epsilon^2 \frac{2 M \vec d}{d^2}\left(\frac{\vec x (t)\!\cdot\! \vec n}{r(t)}- \frac{\vec x_0\!\cdot\! \vec n}{r_0}\right),
\end{equation}
where $\vec x(t)= \vec x_0+ (t-t_0)\vec n+ {\cal O}(\epsilon^2)$ and $\vec d= \vec x_0- (\vec x_0\!\cdot\!\vec n)\vec n$ is the vector joining the centre of the earth and the point of closest approach of the unperturbed ray. Here, $\vec n$ is the initial propagation direction, $\vec n\cdot\vec n=1$.

The leading order post-Newtonian metric is spherically symmetric. Note that the initial polarization that is perpendicular to the propagation plane remains perpendicular to it. So, we select the reference frame differently from that we have done in special relativistic calculations. Here we focus only on gravitational effects and treat them separately from the effects of rotation and relative motion. We take $z=0$, the plane where the ray from GS to SC lies, set their velocities zero and consider the polarization vector
\begin{equation}
    \vec f= (0,0,1)= \mathrm{const.}
\end{equation}

\begin{figure*}[tb]
    \centering
       \includegraphics[width=0.329\textwidth]{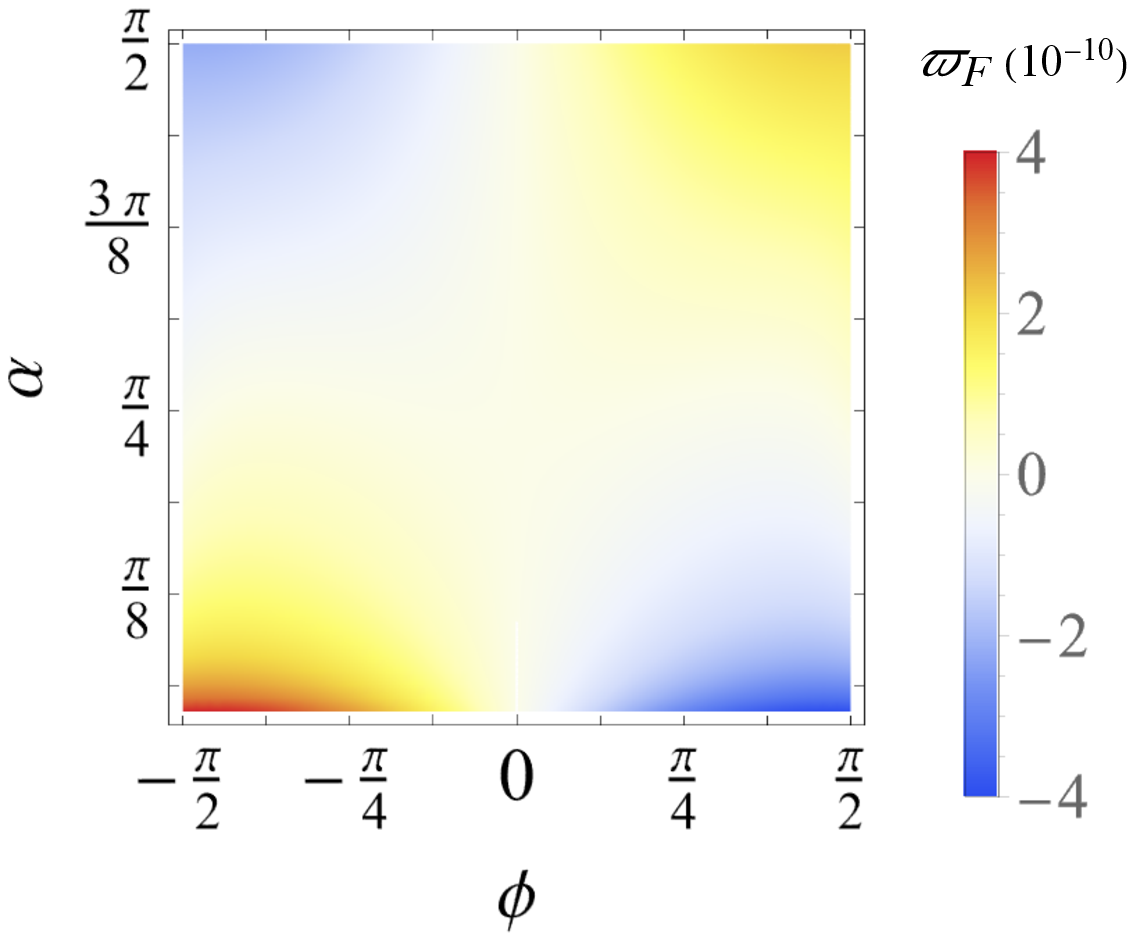}
         \includegraphics[width=0.329\textwidth]{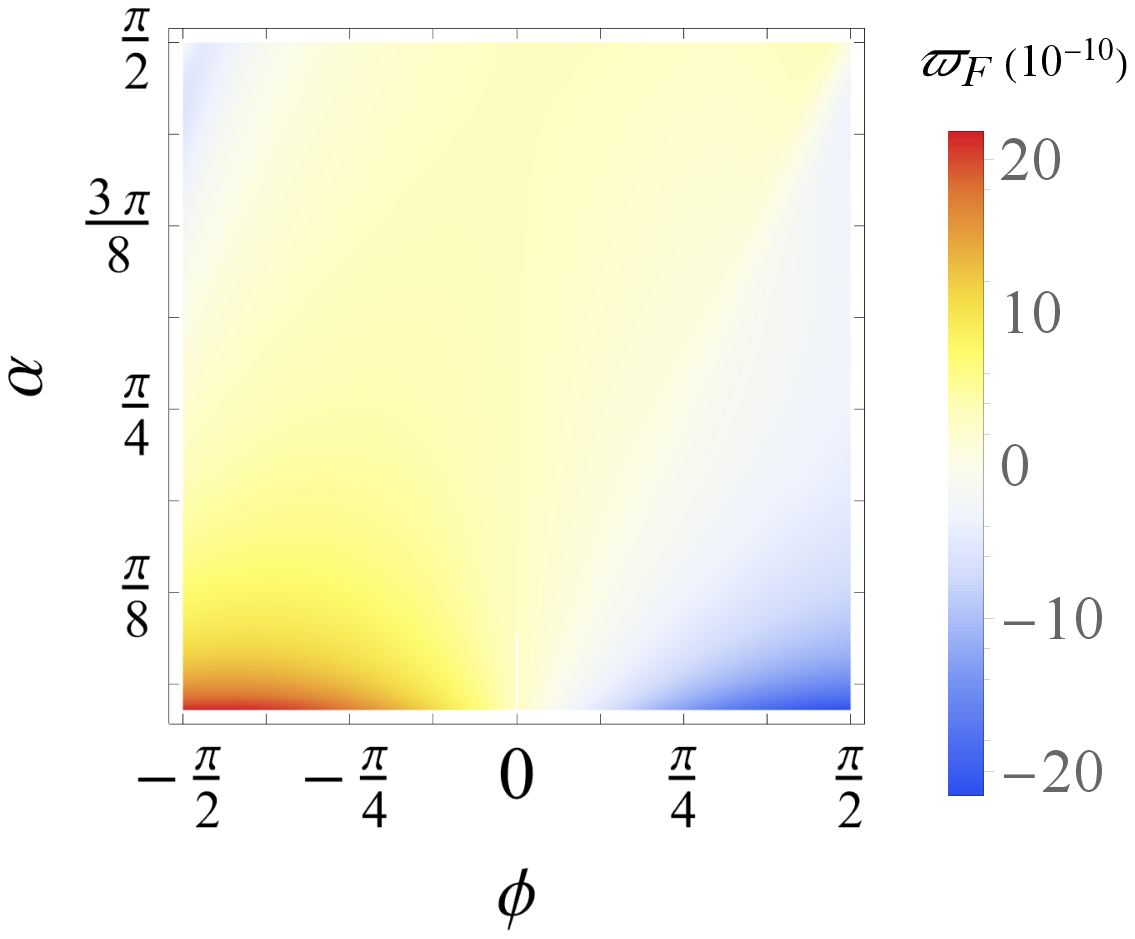}
          \includegraphics[width=0.329\textwidth]{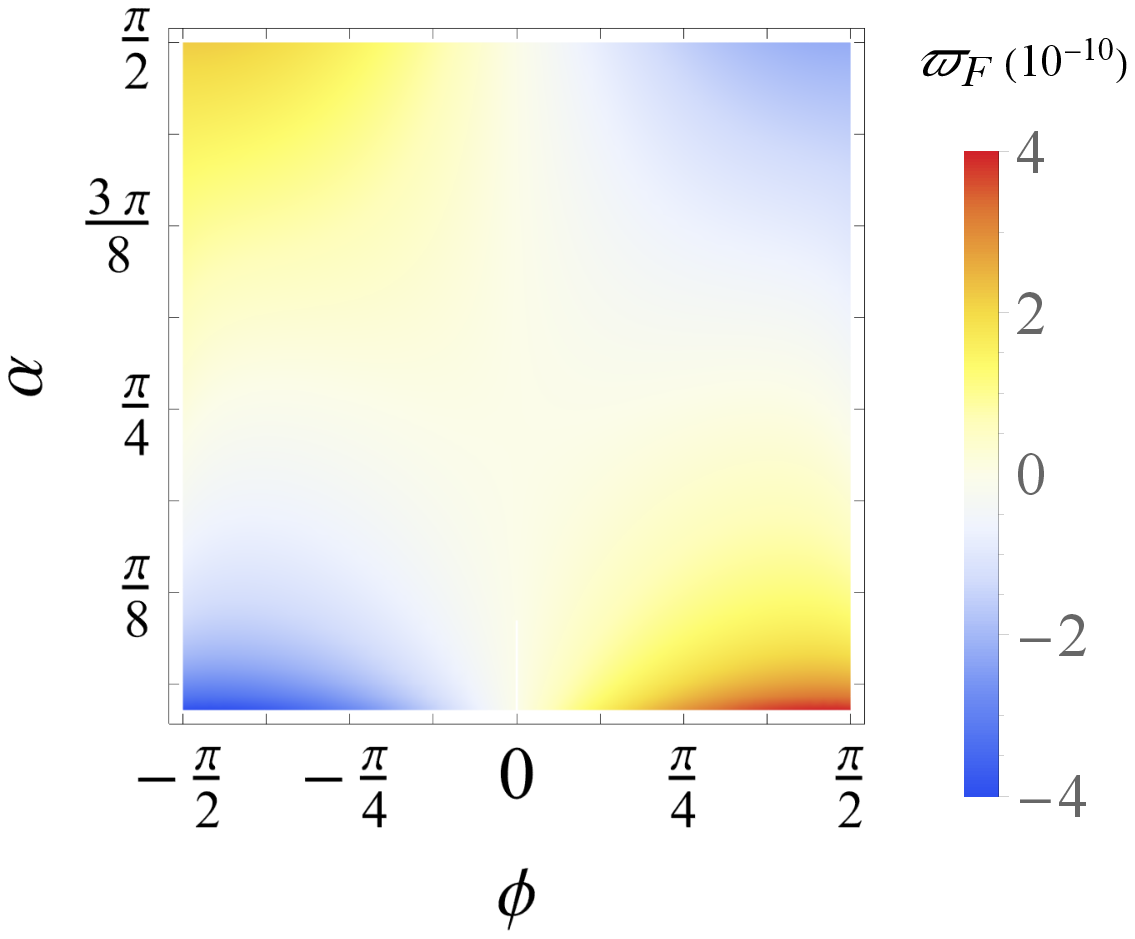}
      \caption{ \label{Faraphase} Faraday phase $\varpi_{\mathrm F}$ depends on the choice of the reference direction $\vec{\zeta}(\alpha,\beta)$ and the initial propagation direction $\vec n(\phi)$. The angles are defined in Eqs.~\eqref{ang29} and \eqref{ang30}. {\it (left)} $\beta=0$. {\it (center)} $\beta=7\pi/16$. {\it (right)} $\beta=\pi$. }
    \label{fig3}
\end{figure*}

The gauge dependent Faraday phase results from the change in the definitions of standard polarization directions along the trajectory. The reference directions $\hzeb_i$ (unit vector pointing to the distant star), $i=$ GS, SC are obtained from the tangents to the rays from the fixed guide star that arrive to the GS and SC respectively. We assume the reference star to be infinitely far. Approximating the differences in reference directions as arising solely form the gravitational field of the earth,
\begin{equation}
    \vec\zeta_i= \vec l\left(1-\epsilon^2 \frac{2 M}{r_i}\right)- \epsilon^2 \frac{2 M \vec d_i}{d_i^2}\left(\frac{\vec x_i\!\cdot\!\vec l}{r_i}-1\right),
\end{equation}
where $\vec l\!\cdot\!\vec l=1$; $-\vec l$ is the flat spacetime direction from the infinitely distant star to the observers. Standard polarization vectors at GS and SC could be defined as
\begin{equation}
    \hbb_{2 i}= \frac{\vec\zeta_i \times \vec k}{|\vec\zeta_i \times \vec k|}, \qquad \hbb_{1 i}= \hbb_{2 i} \times \hbk.
\end{equation}
So, from Eq.~\eqref{ph19}, we get
\begin{equation}
    \Delta\chi\cos\chi= \hbf_{SC}\!\cdot\!\hbb_{2 SC}-\hbf_{GS}\!\cdot\!\hbb_{2GS}
\end{equation}
or,
\begin{equation}
    \Delta\chi= \frac{1}{\hbf_{GS}\!\cdot\!\hbb_{1 GS}}(\hbf_{SC}\!\cdot\!\hbb_{2 SC}-\hbf_{GS}\!\cdot\!\hbb_{2GS}).
\end{equation}

If we choose the x-axis to pass through GS, then
\begin{align}
    &\vec n= (\cos\phi,\sin\phi,0), \qquad -\frac{1}{2}\pi < \phi < \frac{1}{2}\pi,\label{ang29}\\
    & \vec l= (\cos\alpha, \sin\alpha \sin\beta, \sin\alpha \cos\beta), \qquad 0 \leq \alpha \leq \frac{1}{2}\pi,\label{ang30}
\end{align}
where the altitude of the guided star is $\pi/2- \alpha$. When the reference direction $\vec\zeta$ and the propagation direction $\vec k$ are collinear, then the standard polarization directions are undefined. If $\vec\zeta$ lies in the plane determined by GS, SC and centre of earth, then the Faraday phase is zero. Moreover, the post-Newtonian phase fails in the limit of $\alpha=0$ or $\beta= \pi/2$, $3\pi/2$. The plot of the Faraday phase for different reference and propagation direction is shown in Fig.~\ref{fig3}.

\section{Polarization-dependent trajectory}\label{post-G}
Taking into account the post-eikonal terms modifies the description of light propagation along the null rays that are geodesic trajectories of massless particles. These particles are still null, but their motion is no more geodesic. It is most conveniently described by using a specially designed null tetrad\cite{C:92,exact:03}. It is introduced as follows.

Trajectory $X(s)$ of a null particle  with the tangent vector $K$ can be written as
\be
\frac{dX^\mu}{ds}=K^\mu, \qquad K^\mu K_\mu=0.
\ee
This trajectory is typically is not a geodesic\cite{OJDRPA:20,F:20}, and the acceleration vector $\aw$ is defined by
\be
w^\mu\defeq K^\nu K^\mu_{;\nu}, \qquad K^\mu w_\mu=0.
\ee
The second tetrad vector $\an$, $\an^2=0$ is chosen to satisfy
\be
\aK\cdot\an=-1.
\ee
A pair of complex conjugated null vectors, $\am$ and $\bar\am=\am^*$, are built from two spacelike vectors and satisfy
\be
\am\cdot\bar{\am}=1.
\ee
All other inner products vanish and the metric can be expressed as
\be
\sg_{\mu\nu}=-K_{(\mu}n_{\nu)}+m_{(\mu}\bar m_{\nu)}.
\ee
It is possible to reparameterize the null curves and rescale the vectors of the null tetrad while preserving orthonormality relations\cite{C:92,exact:03},
\begin{align}
&\aK\to A\aK, \qquad \an\to A^{-1}\an, \\
&\aK\to \aK, \quad \am\to \am+ a\aK, \quad \bar{\am}\to\bar{\am}+a^*\aK, \quad \an\to \an+a^*\am+a\bar{\am}+a a^*\aK, \\
& \am\to e^{i\phi}\am, \qquad \am\to e^{-i\phi}\bar{\am}.
\end{align}

Using this freedom it is possible to choose the tetrad in such a way\cite{F:20} that the vector $\an$ is parallel propagated along the ray,
\be
 \nabla_\aK \an=0,
 \ee
  while
\be
\nabla_\aK \am=-\kappa \an, \qquad \nabla_\aK \bar\am=-\kappa^* \an,
\ee
where
\be
\kappa =-\aw\cdot\am=-m^\mu K^\nu K_{\mu;\nu}.
\ee
With this choice of the null tetrad the polarization-dependent correction to the trajectory that takes into account the frequency $\omega$ is finite, and given by\cite{F:20}
\be
\frac{D^2X^\mu}{Ds^2}=K^\nu K^\mu_{;\nu}=-\frac{i\sigma}{\omega}R^\mu_{~\nu\alpha\beta}K^\nu m^\alpha\bar{m}^\beta\eqdef F^\mu, \label{om-1}
\ee
where $R^\mu_{~\nu\alpha\beta}$ is the Riemann tensor. At this stage no assumptions about the spacetime were made. Effects of this term in the solar system are adequately described by  the weak field expansion of the Schwarzschild metric. We will shortly see that in the leading order the polarization-dependent acceleration is of the order of $\epsilon^2$. We also treat Eq.~\eqref{om-1} as a correction  to the null geodesic. Thus the uperturbed motion can be confined to the equatorial plane.

The trajectory is decomposed as
\be
X^\mu(s)=x^\mu(s)+\alpha x_\sigma^\mu(s),
\ee
where $x^\mu(s)$ is a null geodesic (that we assume can be affinely parameterized by $s$), $ x_\sigma^\mu(s)$ is the polarization-dependent correction, and   the formal arbitrary infinitesimal parameter $\alpha$ is introduced to ease the perturbative manipulations.  The tangent vector
\be
K^\mu=k^\mu+\alpha \Lambda^\mu,
\ee
 is null so,
\be
k^\mu\Lambda_\mu=0. \label{kLa}
\ee
For the weak gravity regime it is convenient use
the usual post-Newtonian expansion (the standard choice of the coordinates $(t,x,y,z)$, but without switching to the coordinate time $t$ as an evolution parameter). On the dimensional grounds we expect as the
famous light deflection\cite{mtw,will} is the $\epsilon^2$ effect ($\Delta\phi=2r_\sg/\ell$, where $\ell$ is the impact parameter and $r_\sg=2M $is the gravitational radius), the leading polarization-dependent effect is of the order $\alpha=\epsilon^2\lambda/\ell$, where $\lambda$ is the characteristic wave length. Taking this into account and writing $M$ explicitly,
we note that at the leading order
\begin{align}
k^\mu=& k_{(0)}^\mu+Mk_{(1)}^\mu, \\
\Lambda^\mu=& M q^\mu,
\end{align}
where the components $k_{(0)}^\mu$ are constant. Taking into account that $q^\mu_{;\nu}=q^\mu_{,\nu}+\cO(M)$, we have
\be
K^\nu K^\mu_{;\nu}=M k_{(0)}^\nu q^\mu_{,\nu}+\cO(M^2).
\ee

It is convenient to describe the unperturbed null geodesic using the full Schwarzschild solution. The symmetry allows to restrict the motion to the equatorial plane. The tangent can be taken as
\be
k^\mu=\left(\frac{1}{f(\bar{r})},\pm\sqrt{1-\frac{\ell^2 }{\bar{r}^2}f(\bar{r})},0,\frac{\ell}{\bar{r}^2}\right),
\ee
where $f(\bar{r})=1-2M/r$. Here $\bar{r}$ is the Schwarzschild radial coordinate (i.e. the circumferential radius). In the weak field regime it  is related to the isotropic radial coordinate $r=\sqrt{\vx\cdot\vx}$
as $\bar{r}\approx r+M$.

We used the freedom of rescaling the null vectors to set the energy of a fictitious photon to $E=1$, so the reduced angular momentum $\ell\defeq L/E$ equals to the impact parameter\cite{C:92}. The sign of $k^r$ depends on whether the photon approaches the origin (i.e. the Sun),  or moves away from it. In the equatorial plane we have $\ak\nabla_\ak \ak=0$.

The second vector of the null  tetrad
\be
 n^\mu=\left(\frac{\bar{r}^2}{2\big(\bar{r}^2-\ell^2f(\bar{r})\big)},\mp \frac{\bar{r}f(\bar{r})}{2 \sqrt{\bar{r}^2-\ell^2f(\bar{r})}},0,\frac{\ell f(\bar{r})}{2\big(\bar{r}^2-\ell^2f(\bar{r})}\right),
\ee
is chosen so that $\ak\cdot \an=-1$, and the complex-conjugate pair of null vectors is given by
\be
m^\mu=\frac{1}{\sqrt{2}}\left(\frac{i\ell}{\sqrt{\bar{r}^2-\ell^2f(\bar{r})}},0,\frac{1}{r},\frac{i}{\sqrt{\bar{r}^2-\ell^2f(\bar{r})}}\right), \qquad \bar m^\mu=m^{\mu*}.
\ee
These three vectors are to be adjusted   to satisfy the conditions required for the validity of Eq.~\eqref{om-1}. This adjustment is quite straightforward, as in our setting
\be
\nabla_\ak\an=i\alpha(\am-\bar{\am}),
\ee
and
\be
\nabla_\ak\am=i\alpha\ak, \qquad \nabla_\ak\bar{\am}=-i\alpha\ak,
\ee
where
\be
\alpha=\pm\frac{\ell(r-3M)}{\sqrt{2}r(\bar{r}^2-\ell^2f(\bar{r}))},
\ee
where the sign of $\alpha$ coincides with that of $k^r$.

In the leading order approximation the trajectories are geodesic, and
the parallel transport of $\ak$ results in $\kappa=0$. In turn it enables  the parallel transport of the tetrad if the function $a$ satisfies
\be
a_{,r}=-i\alpha/k^r=  \frac{i\ell(\bar{r}-3M)}{\sqrt{2}\big(\bar{r}^2-\ell^2f(\bar{r})\big)^{3/2}},
\ee
where the sign does not change on transition from decreasing to increasing distance.

Null geodesics in the Schwarzschild spacetime are classified \cite{C:92} according to the behaviour of the three roots of the equation $C(\bar{r})=r^3-\ell^2(\bar{r}-2M)=0$. Only one root is relevant for the trajectories we study,
\be
\bar{r}_0=\ell-M+\cO(M^2).
\ee
Up to the terms of the third order in $M$ the roots are $2M$  and $r\pm \bar{r}_0$.
Using this we find
\be
\int a_{,r} dr=\frac{i\ell(3 r M-\bar{r}_0^2)}{\sqrt{2(\bar{r}^2-\bar{r}_0^2)}\bar{r}_0^2}+\cO(M^3),
\ee
that is sufficient to obtain the desired $a(\bar{r})$ at any point of the trajectory apart from $r=\bar{r}_0$.

The same expression is valid  for both increasing and decreasing $r$. For $r>\ell\sim \bar{r}_0\gg M$ we can approximate it as
\be
a=-\frac{i\ell}{\sqrt{2}\sqrt{\bar{r}^2-\ell^2}}-\frac{3iM}{\sqrt{2}\ell}\left(1-\frac{\bar{r}^3}{(\bar{r}^2-\ell^2)^{3/2}}\right)+\cO(M^2).
\ee
It is convenient to express the acceleration $F^\mu$ in terms of the original tetrad vectors as
\be
F^\mu=-\frac{i\sigma}{\omega}R^\mu_{~\alpha\beta\gamma}(k^\alpha m^\beta\bar{m}^\gamma+a k^\alpha k^\beta\bar{m}^\gamma+a^*k^\alpha m^\beta k^\gamma).
\ee
Only one component  is non-zero at the leading order:
\be
F^\theta=-\frac{\sigma}{\omega}\frac{3\ell M}{\bar{r}^4}\left( \frac{1}{\sqrt{\bar{r}^2-\ell^2f(\bar{r})}} +\frac{ \sqrt{2}a(\bar{r})\ell}{\bar{r}^2}\right).
\ee
Using again the decomposition of $C(\bar{r})$ it becomes
\be
F^\theta\approx\frac{\sigma}{\omega} \frac{3\ell M}{\bar{r}^4\sqrt{\bar{r}^2-\bar{r}_0^2}}\left(-1+\frac{\ell^2}{\bar{r}^2}\right)=
-\frac{\sigma}{\omega} \frac{3\ell M}{\bar{r}^6}\frac{{\bar{r}^2-\ell^2}}{\sqrt{\bar{r}^2-\bar{r}_0^2}}  ,\label{a65}
\ee


We evaluate the correction $x_\sigma(s)$  using the post-Newtonian formalism that was outlined above. It is important to note that to use only the leading order  $\epsilon^2$  while using the order of $\alpha$ polarization correction to describe a particular trajectory  is unjustified, as the next post-Newtonian correction for the geometric optics term is still higher. Using the setting that is described below $\epsilon^2=r_\sg/\ell=4.2 \times 10^{-6}$, while $\lambda/\ell=1.4\times 10^{-9}$ for $\lambda=1$m. However, as the exact general relativistic calculation within  the geometric optics approximation leads to polarization-independent results, calculation of the polarization-dependent shift can be done using only the $\epsilon^2$ expansion.

To obtain an estimate of the effect we  neglect the difference between $\bar{r}_0$ and $\ell$, essentially regularizing the expressions near the closest approach  radius.
 The difference between the isotropic radial coordinate that is used in the post-Newtonian analysis and the Schwarzschild radial coordinate is of the order of $2M\ll \ell$ and affects only the higher-order terms.   The unperturbed motion is confined to the $z$-plane.  We can assume that   $\vec{k}=(-1,0,0)$, that corresponds to $\ell>0$.  In this approximation
\be
F^z=+\frac{3\sigma \ell M}{\omega r^5}\sqrt{{r}^2-\ell^2}+\cO(M^2),
\ee
while all other components are zero at the leading order.   Then the explicit equation for the leading  polarization-dependent correction is
\be
q^z_{,x}=-\frac{3\sigma \ell }{\omega r^5}\sqrt{{r}^2-\ell^2}. \label{ode-shift}
\ee

We are interested in a light ray that comes from afar (``minus infinity''), grazes the sun and is detected at some finite (but large) distance $r_2\gg \ell$, the initial off-plane displacement and velocity are zero.
Similarly to the calculations of corrections for the classical tests of general relativity\cite{mtw,will} we use the unperturbed (i.e. flat spacetime trajectory with $r_0=\ell$
\be
y=\ell, \qquad x=-t.
\ee
Then the correction $\delta z=M q^z$ is obtained by an elementary integration. For the far field ($r\gg \ell$) the trajectory approaches a straight line with
\be
\theta_\infty=\frac{2\sigma}{\omega}\frac{M}{\ell^2},
\ee
giving at the distance $r$ from the sun the transversal shift
\be
\Delta z= \frac{4M}{\omega\ell^2}r.
\ee

This shift coincides with the result obtained using the polarization-dependent part of the angle of deflection
$\Delta\phi$ between in-and outgoing polarized rays that was reported in Ref.~\citenum{GBM:07}. On the other hand, this approximation is too crude to identify a  convergence of the initially divergent trajectories that  was reported in Ref.~\citenum{OJDRPA:20}, and a more delicate approximations will be investigated in the follow-up work.

\section{Summary}

Describing propagation of electromagnetic waves in vacuum  in terms of  rays that follow null geodesics is a very good approximation in the high-frequency regime. Within this approximation  the polarization rotation in the Schwarzschild metric, and as a result in the leading post-Newtonian approximation, is a purely gauge effect. This phase will be present as a consequence of practical methods of setting up reference frames in the Earth-to-spacecraft communications.    However,  for the near-Earth experiments  these effects are typically about $10^{-5}$ weaker than the SR effects.
Observable gravity-induced deviations from the geometric optics approximation are expected only in ultra-strong gravitational fields.

\end{document}